\documentclass[aps,11pt]{article}
\usepackage{epsfig,sint,macros,cite,mathbbol,graphicx,amssymb}

\begin{document}
\begin{titlepage}
\begin{flushright}
MKPH-T-09-13
\end{flushright}

\renewcommand{\thefootnote}{\fnsymbol{footnote}}

\vskip 0.5 cm
\begin{center}
  {\Large\bf Minimally doubled fermions at one loop \\[0.5ex]}
\end{center}
\vskip 0.5 cm
\begin{center}
{\large Stefano Capitani\footnote{capitan@kph.uni-mainz.de}, Johannes
  Weber\footnote{weberj@kph.uni-mainz.de} and Hartmut
  Wittig\footnote{wittig@kph.uni-mainz.de}}
\vskip 0.5cm
Institut f\"ur Kernphysik, Becher Weg 45,\\
University of Mainz, D-55099 Mainz, Germany
\vskip 1.0cm
{\bf Abstract}
\vskip 0.35ex
\end{center}

\renewcommand{\thefootnote}{\arabic{footnote}}

\noindent
Minimally doubled fermions have been proposed as a cost-effective
realization of chiral symmetry at non-zero lattice spacing. Using
lattice perturbation theory at one loop, we study their
renormalization properties. Specifically, we investigate the
consequences of the breaking of hyper-cubic symmetry, which is a
typical feature of this class of fermionic discretizations. Our
results for the quark self-energy indicate that the four-momentum
undergoes a renormalization which is linearly divergent. 
We also compute renormalization factors for quark bilinears, construct the
conserved vector and axial-vector currents and verify that at one loop
the renormalization factors of the latter are equal to one.

\vfill

\begin{center}
August 2009
\end{center}

\eject

\vfill
\eject

\end{titlepage}

\setcounter{footnote}{0}

\par\noindent
{\bf 1.} The past ten years have witnessed two major breakthroughs in
lattice QCD. The first concerns the significant acceleration of
simulation algorithms for dynamical quarks, in particular as the light
quark masses are tuned towards the chiral regime. Owing to these
developments, simulations with pion masses close to the physical value
have become routine. The second achievement was the solution to the
long-standing problem of constructing discretizations of the quark
action which preserve chiral symmetry, and the realization of the
r\^ole of the Ginsparg-Wilson relation\,\cite{GWR,ML_GWR98}. However,
the well-known discretizations based on the perfect action formalism
\cite{Hasenfratz:1997ft}, or, alternatively, the domain wall
\cite{dwf:Kaplan,dwf:FurSha} and overlap constructions
\cite{ovlp:Neuberger} all involve non-local interactions. As a
result, their implementation is numerically much more expensive than
conventional Wilson \cite{Wilson74} or staggered
\cite{staggered} fermions. 

Minimally doubled fermions
\cite{mind:Karsten81,mind:Wilczek87,mind:Creutz07,mind:Borici07,mind:Creutz08}
share the desirable features of strict locality with traditional
discretizations, whilst preserving exact chiral symmetry for a
degenerate doublet of quark fields. The question whether or not they
are suitable for the determination of hadronic properties in practical
simulations has not been thoroughly investigated so far. Before
embarking on extensive numerical studies of minimally doubled
fermions, it is useful to examine some of their properties in
perturbation theory. In this letter, we present our results for the
quark self-energy and the renormalization properties of quark
bilinears at one loop in lattice perturbation theory. In particular,
we shall elucidate the consequences of the breaking of hyper-cubic
symmetry, which is a typical feature of this class of lattice
actions. Our main findings indicate that hyper-cubic symmetry breaking
generates a renormalization of the quark's four-momentum.

After fixing our notation in section~2, we list expression for the
quark propagator and the vertices in section~3. The perturbative
calculation of the quark self-energy at one loop in lattice
perturbation theory is presented in section~4. Sections~5 and~6
discuss the renormalization properties of quark bilinears and provide
expressions for the conserved vector and axial-vector currents. In
section~7 we present our conclusions and discuss the consequences of
our results for numerical simulations.

\bigskip
\bigskip

\par\noindent
{\bf 2.} Following the works of Bori\c{c}i\,\cite{mind:Borici07} and
Creutz\,\cite{mind:Creutz08} we employ the particular construction of
minimally doubled fermions of ref.\,\cite{mind:Creutz08}. The
corresponding lattice Dirac operator respects chiral symmetry, and is
also $\rmO(a)$-improved. For massless quarks, the general expression
reads
\be
  {\cal{D}} = D+\overline{D}-2i\Gamma.
\label{eq:creutz-action}
\ee
In momentum space the terms $D$ and $\overline{D}$ are given by
\be
          D(p) = i \, \sum_\mu (\gamma_\mu \sin p_\mu),\qquad
\overline{D}(p)= i \, \sum_\mu (\gamma'_\mu \cos p_\mu).
\ee
The matrices $\Gamma$ and $\gamma'_\mu$ are defined by
\be
  \Gamma = \frac{1}{2} \, \sum_{\mu}\gamma_\mu,
  \qquad\gamma'_\mu = \Gamma \gamma_\mu \Gamma = \Gamma - \gamma_\mu ,
\label{eq:Gamma_def}
\ee
with $\Gamma^2=1$. Other useful relations are
\be
  \sum_\mu \gamma_\mu = \sum_\mu \gamma'_\mu = 2 \Gamma , \quad
  \{ \Gamma, \gamma_\mu \} = 1 , \quad
  \{ \Gamma, \gamma'_\mu \} = 1 .
\ee
The construction of the Dirac operator ${\cal{D}}$ involves a
particular linear combination of two (physically equivalent)
na\"{\i}ve fermion actions, corresponding to $D$ and $\overline{D}$
in\,\eq{eq:creutz-action}. The first term, as is widely known, has~16
zeros in the first Brillouin zone, when any component of $p$ is equal
to~$0$ or~$\pi$. The term $\overline{D}$ has also 16 zeros, which are,
however, positioned at the points where $p_\mu=\pm\pi/2$.

As was shown in ref.\,\cite{mind:Creutz08}, the presence of the term
$-2i\Gamma$ in \eq{eq:creutz-action} guarantees that ${\cal{D}}(p)$
exhibits only two Fermi points, located at $p=(0,0,0,0)$ and
$p=(\pi/2,\pi/2,\pi/2,\pi/2)$, which represent two degenerate fermion
species of opposite chirality. Here we simply note that the extra
zeros of $D$ at the corners of the Brillouin zone are lifted by the
presence of $\overline{D}-2i\Gamma$. An entirely similar statement
applies to the zeros of $\overline{D}$.

The matrix $\Gamma$ is not unique: There are altogether 16 definitions
of $\Gamma$, in which the coefficient of a particular gamma-matrix can
be chosen as $1$ or $-1$. Each choice selects different zeros of
${\cal{D}}$, which correspond to the physical degrees of freedom, but
otherwise all such definitions yield an equivalent theory.

The inclusion of the term proportional to $\Gamma$ implies that the
action is no longer symmetric under the full hyper-cubic
group. Depending on its definition, the matrix $\Gamma$ selects a
particular direction in Euclidean space. The action of minimally
doubled fermions is only symmetric with respect to a subgroup of the
full hyper-cubic group, which preserves this fixed direction (up to a
sign). For the action considered in this paper, which corresponds to
the definition of $\Gamma$ in \eq{eq:Gamma_def}, this is the positive
major diagonal. For other minimally doubled actions, such as those
considered in refs.\,\cite{mind:Karsten81,mind:Wilczek87}, it is the
temporal axis.

The breaking of hyper-cubic symmetry implies the possibility of mixing
with operators of different dimensionality, such as
$\overline{\psi}\Gamma\Box\psi$. In
refs.\,\cite{Bedaque:2008xs,Bedaque:2008jm,Buchoff:2008ei} it was
argued that mixing with dimension-3 operators cannot be avoided. In
any case, the lack of hyper-cubic symmetry generates not only mixing
with dimension-3 operators, but also mixing with marginal operators of
dimension~4. In the next two sections we will show that hyper-cubic
symmetry breaking generates a linearly divergent additive
renormalization of the quark's momentum.

\bigskip
\bigskip

\par\noindent
{\bf 3.} By inverting the Dirac operator of \eq{eq:creutz-action}, and
restoring the proper factors of $a$, we obtain the fermion propagator
\begin{equation}
S(p) = a\,\frac{\displaystyle -i\sum_\mu \gamma_\mu
       (\sin ap_\mu-\cos ap_\mu)  
      -i\Gamma (\sum_\mu \cos ap_\mu -2) +am_0}{\displaystyle
   \sum_\mu \Big[\sin ap_\mu \sum_\nu \cos ap_\nu -2 \sin ap_\mu (\cos
   ap_\mu+1) -2\cos ap_\mu\Big] +8  +(am_0)^2} .
\end{equation}
Unlike many standard fermionic discretizations one finds that the
denominator of this propagator cannot be cast into a form which
possesses a definite behaviour under parity transformation in each
single coordinate ($p_i \to -p_i$). This is not surprising in view of
the fact that the definition of $\Gamma$ singles out an intrinsic
Euclidean direction.

By using the identities
$\{\gamma_\mu,\gamma_\nu\}=\{\gamma'_\mu,\gamma'_\nu\}=2\delta_{\mu\nu}$
and $\{\gamma_\mu,\gamma'_\nu\}=1-2\delta_{\mu\nu}$, the above
propagator can also be written in a form which is more convenient for
lattice perturbation theory, i.e.
\begin{equation}
S(p) = a\,\frac{\displaystyle -i\sum_\mu \Big[ \gamma_\mu \sin ap_\mu 
            -2\,\gamma'_\mu\,\sin^2 ap_\mu/2 \Big] +am_0}{\displaystyle 
   4 \sum_\mu \Big[ \sin^2 ap_\mu/2 + \sin ap_\mu 
      \Big(\sin^2 ap_\mu/2 
      -\half \sum_\nu \sin^2 ap_\nu/2 \Big)\Big]
      +(am_0)^2},
\label{eq:propFP0}
\end{equation}
where the limit of small $p$ (i.e. the continuum limit) becomes more
transparent. In particular, one can use the standard methods for the
treatment of lattice divergences, and calculate for example the
so-called $J$ and $I-J$ parts along the lines of \cite{Kawai:1980ja}.

Via the substitution $ap_\mu \to \pi/2 + ap_\mu$ one obtains the
propagator for the fermionic mode associated with the Fermi point at
$ap_\mu=\pi/2$:
\begin{equation}
S'(p) = a\,\frac{\displaystyle -i\sum_\mu \Big[ -\gamma'_\mu \sin ap_\mu 
            -2\,\gamma_\mu\,\sin^2 ap_\mu/2 \Big] +am_0}{\displaystyle 
   4 \sum_\mu \Big[ \sin^2 ap_\mu/2 - \sin ap_\mu 
      \Big(\sin^2 ap_\mu/2 
      -\half \sum_\nu \sin^2 ap_\nu/2 \Big)\Big]
      +(am_0)^2} .
\end{equation}
By changing the direction of the four-momentum $p_\mu$ and exchanging
$\gamma_\mu$ with $\gamma'_\mu$, one recovers the propagator of
\eq{eq:propFP0}. Since $\gamma'_5=-\gamma_5$ this implies that the
modes corresponding to the two Fermi points have indeed opposite
chirality.

The quark-quark-gluon vertex is derived as
\begin{equation}
V_1(p_1,p_2) = - i g_0 \left(\gamma_\mu \cos \frac{a(p_1+p_2)_\mu}{2}
       -\gamma'_\mu \sin \frac{a(p_1+p_2)_\mu}{2} \right) ,
\end{equation}
and the quark-quark-gluon-gluon vertex is
\begin{equation}
V_2(p_1,p_2) = \frac{1}{2} i a g_0^2 \left(\gamma_\mu \sin 
     \frac{a(p_1+p_2)_\mu}{2} +\gamma'_\mu \cos
     \frac{a(p_1+p_2)_\mu}{2} \right) , 
\end{equation}
where $p_1$ and $p_2$ are the incoming and outgoing quark momenta at
the vertex. Following ref.\,\cite{mind:Borici08}, the expressions for
the vertices can be easily derived by comparing the Dirac operator of
minimally doubled fermions,
\eq{eq:creutz-action}, with the Wilson-Dirac operator
\begin{equation}
D_{\rm w}(p) = \frac{1}{a}\sum_\mu \Big\{i\gamma_\mu \sin ap_\mu + r \,
(1-\cos ap_\mu)\Big\} + m_0, 
\label{eq:wilson-action}
\end{equation}
and noting that the hopping terms of these two actions are related by
the replacement $r \rightarrow -i\gamma'_\mu$. Indeed, since the terms
$e^{iap_\mu}$ and $e^{-iap_\mu}$ are coupled to the Fourier transforms
of $U_\mu(x)$ and $U_\mu^\dagger(x-a\hat{\mu})$ respectively, it is
sufficient to substitute $r \to -i\gamma'_\mu$, in order to obtain the
vertices for minimally doubled fermions from those of the Wilson case.

\bigskip
\bigskip

\par
\begin{figure}[t]
\begin{center}
\includegraphics[width=9.cm]{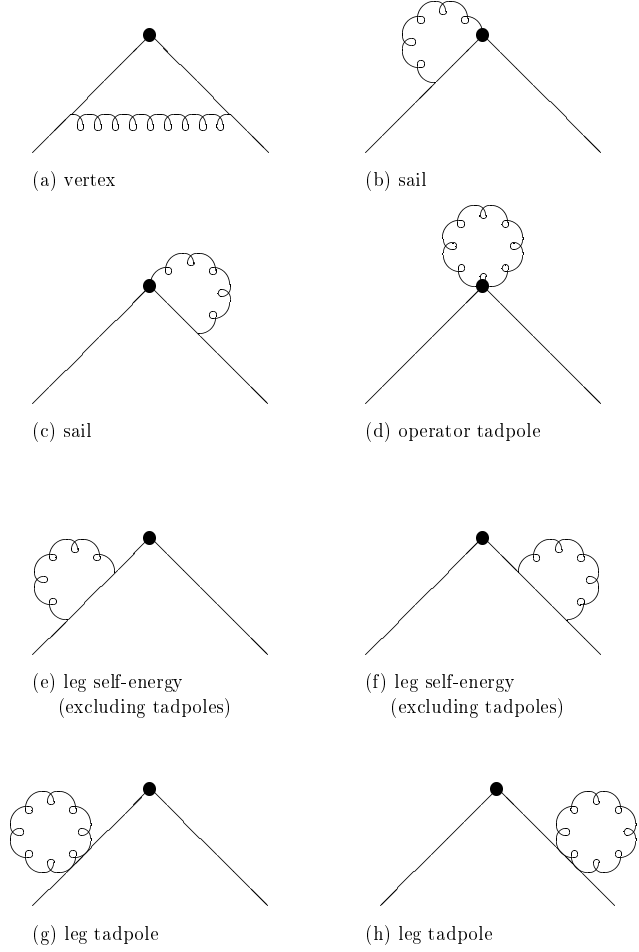}
\caption{\label{fig:diagrams}\small The diagrams needed for the one-loop
renormalization of the lattice operators.}
\end{center}
\end{figure}

\noindent
{\bf 4.} We are now going to describe the calculation of the quark
self-energy at one loop. Figure\,\ref{fig:diagrams} lists all diagrams
which are relevant for the perturbative calculation in this letter.

Using the expression for the vertex $V_2(p,p)$, the tadpole
contribution to the self energy (diagrams (g) and (h) in
Fig. \ref{fig:diagrams}) is easily computed. In a general covariant
gauge, where $\partial_\mu A_\mu=0$, the expression is
\bea
& & \frac{1}{a^2} \cdot \frac{Z_0}{2} \Big(1-\frac{1}{4}(1-\alpha)\Big) 
\cdot i a g_0^2 C_F \sum_\mu \Big( \gamma_\mu a p_\mu 
+ (\Gamma-\gamma_\mu) (1+\rmO(a^2) \Big)  \nonumber \\
&=& g_0^2 C_F \, \frac{Z_0}{2} \Big(1-\frac{1}{4}(1-\alpha)\Big) 
\, \Big( i\slash{p} + \frac{i}{a} 
\sum_\mu (\Gamma-\gamma_\mu) \Big) +\rmO(a),
\eea
where $Z_0$ is given by
\cite{GonzalezArroyo:1981ce,Ellis:1983af,Capitani:2002mp} 
\be
  Z_0=\int_{-\pi/a}^{\pi/a} \frac{d^4p}{(2\pi)^4}
     \frac{1}{\widehat{p}^2} = 0.1549333\ldots 
     = 24.466100\,\frac{1}{16\pi^2},\qquad
     \widehat{p}^2=\frac{4}{a^2}\,\sum_\mu \sin^2\Big(\frac{ap_\mu}{2}\Big).
\ee
Terms of $\rmO(a)$ and higher are not important here. Since $\sum_\mu
\gamma_\mu = 2\Gamma$, the result of the one-loop tadpole is
\begin{equation}
g_0^2 C_F \frac{Z_0}{2}\Big(1-\frac{1}{4}(1-\alpha)\Big) 
\, \Big( i\slash{p} + \frac{2i\Gamma}{a} \Big) .
\end{equation}
The term proportional to $i\slash{p}$ is the same as for Wilson
fermions, while the other term, as already noted in
\cite{Bedaque:2008xs}, would imply a power-divergent mixing of
order~$1/a$ with the dimension-three operator $\overline{\psi} \Gamma
\psi$, provided that there is no cancellation by an analogous term
coming from the contribution of the sunset diagrams to the self-energy
(diagrams (e) and~(f) in Fig. \ref{fig:diagrams}). In this section we
show that there is no such compensation.

We have computed the sunset diagram using special computer codes
written in FORM \cite{Vermaseren:2000nd,Vermaseren:2008kw} and
Mathematica, and also checked it against calculations by hand. The
result of this diagram is
\begin{eqnarray}
& &\Sigma^{sunset} (p,m_0) =
i\slash{p}\cdot\frac{g_0^2}{16\pi^2} \,C_F \,\Bigg[ \log a^2p^2 -5.42642
+(1-\alpha) \Big(-\log a^2p^2 +7.850272 \Big) \Bigg] \nonumber\\
&&\quad +\,m_0\cdot\frac{g_0^2}{16\pi^2} \,C_F \,\Bigg[ 4\,\log a^2p^2
  -29.48729   
+(1-\alpha) \Big(-\log a^2p^2 +5.792010 \Big) \Bigg]  \\
&&\quad +\,1.52766 \cdot\frac{g_0^2}{16\pi^2} \,C_F \cdot i\, \Gamma
\sum_\mu p_\mu  
 +\,(5.07558 + 6.11653\,(1-\alpha))
\cdot\frac{g_0^2}{16\pi^2} \,C_F \cdot i\, \frac{\Gamma}{a} .\nonumber
\end{eqnarray}
Note that gauge invariance forces the terms proportional to
$(1-\alpha)$ to be the same as, for example, in the case of Wilson or
overlap fermions. This is an important check of the correctness of our
calculations.

The total contribution of the one-loop diagrams to the quark
self-energy is then
\begin{equation}
\Sigma (p,m_0) = i\slash{p}\,\Sigma_1(p) +m_0\,\Sigma_2(p) 
+ c_1 (g_0^2)\cdot i\, \Gamma \sum_\mu p_\mu
+ c_2 (g_0^2)\cdot i\, \frac{\Gamma}{a},
\label{eq:totalself}
\end{equation}
where
\bea
\Sigma_1(p) &=& \frac{g_0^2}{16\pi^2} \,C_F \,\Bigg[ \log a^2p^2
  +6.80663 \nonumber \\
& & \phantom{\frac{g_0^2}{16\pi^2} \,C_F \,\Bigg[}
+(1-\alpha) \Big(-\log a^2p^2 + 4.792010 \Big) \Bigg]
\label{eq:Sigma1self} \\ 
\Sigma_2(p) &=& \frac{g_0^2}{16\pi^2} \,C_F \,\Bigg[ 4\,\log
  a^2p^2 -29.48729 \nonumber \\
& & \phantom{\frac{g_0^2}{16\pi^2} \,C_F \,\Bigg[}
+(1-\alpha) \Big(-\log a^2p^2 +5.792010 \Big) \Bigg]
\label{eq:Sigma2self} \\ 
c_1 (g_0^2)\, &=&  1.52766 \cdot\frac{g_0^2}{16\pi^2} \,C_F
 \label{eq:c1self} \\
c_2 (g_0^2)\, &=& 29.54170 \cdot\frac{g_0^2}{16\pi^2} \,C_F .
\eea
As indicated above, the two terms proportional to $\Gamma/a$ arising
from the tadpole and the sunset diagrams do not cancel --- they
actually reinforce each other. Note, however, that the parts
proportional to $(1-\alpha)$ cancel exactly, as required by gauge
invariance.

The full inverse propagator at one loop can be written as 
\begin{equation}
\Sigma^{-1} (p,m_0) = \Big( 1 -\Sigma_1 -\frac{c_1}{2} \Big) \cdot
\Big\{ i\slash{p}
+ m_0 \,\Big( 1 -\Sigma_2 +\Sigma_1 +\frac{c_1}{2}\Big) 
-\frac{ic_1}{2} \,\sum_{\mu\neq\nu} \gamma_\mu p_\nu
-\frac{ic_2}{a}\,\Gamma \Big\}, 
\end{equation}
where we have collected all terms proportional to $i\slash{p}$ in the
wave-function renormalization, which then contains, in addition to the
standard term $\Sigma_1$, also $c_1$. By contrast, the linear
divergence (i.e. the term proportional to $c_2$) must be absorbed into
a redefinition of the four-momentum, which amounts to a uniform
additive shift,
\begin{equation}
p'_\mu = p_\mu -\frac{c_2(g_0^2)}{2a} .
\label{eq:p-redefinition}
\end{equation}
After replacing $p$ by $p'$ and neglecting terms of $\rmO(g_0^4)$, we obtain
\begin{equation}
\Sigma (p',m_0) = \frac{Z_2}{i\slash{p}' + \zm\, m_0 
-i\half c_1(g_0^2) \sum_{\mu\neq\nu} \gamma_\mu p_\nu},
\label{eq:renprop}
\end{equation}
where the wave-function renormalization at one loop is given by
\begin{equation}
Z_2 = \Big( 1 -\Sigma_1 -\frac{c_1}{2} \Big)^{-1},
\end{equation}
while
\begin{equation}
\zm = 1 - \Big( \Sigma_2 -\Sigma_1 -\frac{c_1}{2} \Big) 
\label{eq:zm}
\end{equation}
is the result for the quark mass renormalization factor.

Our results demonstrate that the self-energy generates a
power-divergent mixing with an operator of the form $\Gamma/a$. As can
be seen from the Dirac structure, this mixing is not a renormalization
of the mass. Indeed, chiral symmetry protects the quark mass against
an additive renormalization like in the Wilson case. Rather, the
power-divergent mixing implies that all components of the
four-momentum $p_\mu$ are shifted under renormalization by an equal
amount, i.e. $p_\mu \to p'_\mu = p_\mu +{\rm const.}/a$. The constant
can be determined either order by order in perturbation theory, or at
the non-perturbative level in a Monte Carlo simulation. For instance,
in our perturbative calculation it is given by $-c_2(g_0^2)/2$.

It is important to realize that the term proportional to $c_1(g_0^2)$
cannot be absorbed into a redefinition of $p_\mu$, since otherwise the
conserved vector and axial-vector currents do not have unit
normalization.  We address this issue in more detail in section~6.
Thus, the renormalized quark propagator, eq.\,(\ref{eq:renprop}),
contains also a term $\sum_{\mu\neq\nu} \gamma_\mu p_\nu$. This should
not come as a surprise, since the present formalism is no longer
isotropic.  Note that the presence of this term does not move the pole
of the propagator at $p=0$.

Since the mass is protected from an additive renormalization, the
redefinition of the four-momentum amounts to a renormalization of the
velocity. Noting that $p_\mu$ is a fermionic momentum (it is the
external momentum of the fermionic self-energy), and that no such
phenomenon can occur for the gluonic self-energy, we can interpret
this mixing as a renormalization of the quark velocity. These findings
support Creutz's conjecture
\cite{mind:Creutz07,mind:Creutz08} that ``interactions at finite
lattice spacing can result in the gluons and fermions not having the
same speed of light.''

We remark that the power-divergent mixing proportional to $c_2$, as
well as the one of the same dimensionality (proportional to $c_1$),
occur among operators which are not invariant under the hyper-cubic
group. Such mixings are lattice artefacts which are peculiar to
minimally doubled fermions.

Once the above subtraction has been made, and $p_\mu$ replaced with
$p'_\mu$ in all renormalized quantities, the power divergence
disappears. Although we have no proof, it is not unreasonable to
expect that this can be done consistently at every order of
perturbation theory, similar to the subtraction of the $1/a$
divergence in the self-energy of Wilson fermions, which is
consistently removed from the theory by the replacement $m_0 \to
m_q=m_0-m_{cr}$.

One may wonder how the redefinition of the four-momentum affects
numerical simulations and how it could be determined
non-perturbatively. We postpone this discussion to our conclusions in
section~7.

\bigskip
\bigskip

\par\noindent
{\bf 5.} We have also computed the renormalization factors of local
bilinears, and we list the results for the vertex diagrams below. The
complete renormalization factors are obtained after including
wave-function renormalization, which is achieved by adding the
contributions of $\Sigma_1$ and $c_1$, eqs.\,(\ref{eq:Sigma1self})
and~(\ref{eq:totalself}), of the self-energy. For the scalar density
the result for the vertex (diagram (a) in Fig.\,\ref{fig:diagrams}) is
\begin{equation}
\frac{g_0^2}{16\pi^2} \,C_F \,\Bigg[ -4\,\log a^2p^2 + 29.48729
+(1-\alpha) \Big(\log a^2p^2 -5.792010 \Big) \Bigg] .
\label{eq:lambda-s}
\end{equation}
Here there is no mixing term arising from the breaking of hyper-cubic
invariance. The only such mixing occurs after adding the wave-function
contribution, which includes the term proportional to $c_1(g_0^2)$. 
For the vector current the vertex diagram yields
\begin{equation}
\frac{g_0^2}{16\pi^2} \,C_F \,\gamma_\mu \Bigg[ -\log a^2p^2 + 9.54612
+(1-\alpha) \Big(\log a^2p^2 -4.792010 \Big) \Bigg] 
+c_{\rm v}^{\rm vtx}(g_0^2) \cdot \Gamma,
\end{equation}
where the coefficient of the mixing is given by
\begin{equation}
c_{\rm v}^{\rm vtx}(g_0^2)\, =  -0.10037 \cdot
\frac{g_0^2}{16\pi^2} \,C_F . 
\end{equation}
This is a mixing with an operator of the same dimension, which is not
invariant under the hyper-cubic group. Note that there can be no
power-divergent mixing here (and in all the other bilinears), as one
can see by simple dimensional counting.

As a consequence of chiral symmetry, the vertex corrections are
identical for the vector and axial-vector currents, and the same is
true also for the scalar and pseudo-scalar densities. We have verified
this in the course of our calculations. After taking the
renormalization of the wave-function into account, the renormalization
factors $\zv$ and $\za$ of the local vector and axial-vector currents
are not equal to one. In order to identify the conserved currents,
which are protected against renormalization, one has to consider the
chiral Ward identities. We postpone this discussion to the next
section.

Finally, for the tensor current we obtain the result for the vertex
diagram as
\begin{equation}
\frac{g_0^2}{16\pi^2} \,C_F \,\sigma_{\mu\nu}\,\Bigg[ 2.16548
+(1-\alpha) \Big(\log a^2p^2 -3.792010 \Big) \Bigg] .
\end{equation}
Again, the breaking of hyper-cubic invariance does not generate any
extra mixing, apart from the one arising from the self-energy.

We end this section with a brief comment on the renormalization of the
quark mass. Chiral symmetry protects the bare quark mass $m_0$ from
undergoing an additive renormalization. The relation between the bare
and renormalized quark masses, $m_0$ and $m_{\rm R}$, respectively, is
then obtained via
\be
   m_{\rm R} = \zm\,m_0,
\ee
where $\zm$ is given in eq.\,(\ref{eq:zm}). The full expression for
the renormalization factors of the scalar and pseudo-scalar densities
in perturbation theory at one loop is
\be
   \zs = \zp = 1 - \Big( \Lambda_{\rm S}+\Sigma_1+\frac{c_1}{2} \Big),
\ee
where the results for the self-energy contributions $\Sigma_1,$ and
$c_1$ are given in eqs.\,(\ref{eq:Sigma1self}) and\,(\ref{eq:c1self}).
Here $\Lambda_{\rm S}$ is the result for the one-loop vertex diagram
of the scalar density, given in eq.\,(\ref{eq:lambda-s}), which is
exactly equal to the $\rmO(g_0^2)$-contribution to the quark
self-energy $\Sigma_2$, but comes with an opposite sign. Thus, when we
compare with eq.\,(\ref{eq:zm}), we see that the renormalization
factors $\zs$ and $\zp$ of the scalar and pseudo-scalar densities
satisfy
\be
    1/\zm=\zs=\zp,
\ee
where the last equality is a consequence of chiral symmetry. We have
thus verified at one loop in perturbation theory, that the
renormalization of the quark mass for minimally doubled fermions has
the same form as, say, in the case of overlap fermions.

\bigskip
\bigskip

\par\noindent
{\bf 6.} One can derive expressions for the conserved vector and
axial-vector currents via the chiral Ward identities along the lines
of ref.\,\cite{Boch}. The lattice action of minimally doubled fermions
in position space reads
\begin{eqnarray}
& & S^f = a^4 \sum_{x} \bigg[ \frac{1}{2a} \sum_{\mu} \Big[
    \overline{\psi} (x)  
(\gamma_\mu + i\gamma'_\mu) U_\mu (x) \psi (x + a\widehat{\mu}) 
\nonumber \\
&& \phantom{S^f = a^4 \sum_{x} \bigg[ \frac{1}{2a} \sum_{\mu}}
-\overline{\psi} (x + a\widehat{\mu}) (\gamma_\mu - i\gamma'_\mu)
U_\mu^\dagger (x) \psi (x) \Big] +
\overline{\psi}(x)\Big(m_0-\frac{2i\Gamma}{a}\Big) \psi (x) 
    \bigg].
\end{eqnarray}
It is important to recall that the Dirac spinor $\psi$ in this
expression describes a degenerate doublet of quarks. The action in the
massless case is invariant under an axial U(1) transformation, and it
is then clear that the chiral Ward identities associated with this
exact symmetry yield the isospin-singlet currents of the theory.

If one applies the usual vector and axial transformations, i.e.
\bea
& &\delta_V \psi = i \alpha_V \psi , \qquad\phantom{\gamma_5}
   \delta_V \overline{\psi} = - i \overline{\psi} \alpha_V , \nonumber \\
& &\delta_A \psi = i \alpha_A \gamma_5 \psi , \qquad
   \delta_A \overline{\psi} = i \overline{\psi} \alpha_A \gamma_5 ,
\eea
one identifies the conserved vector current as
\begin{equation}
V_\mu^{\mathrm c} (x) = \frac{1}{2} \bigg(
   \overline{\psi} (x) \, (\gamma_\mu+i\,\gamma'_\mu) \, U_\mu (x) \, 
   \psi (x+a\widehat{\mu}) 
 + \overline{\psi} (x+a\widehat{\mu}) \, (\gamma_\mu-i\,\gamma'_\mu) \, 
   U_\mu^\dagger (x) \, \psi (x) \bigg) ,
\label{eq:noether-vector}
\end{equation}
while the axial-vector current (which is conserved in the massless
case) is given by
\begin{equation}
A_\mu^{\mathrm c} (x) = \frac{1}{2} \bigg(
   \overline{\psi} (x) \, (\gamma_\mu+i\,\gamma'_\mu) \, \gamma_5 \,
   U_\mu (x) \, 
   \psi (x+a\widehat{\mu}) 
 + \overline{\psi} (x+a\widehat{\mu}) \, (\gamma_\mu-i\,\gamma'_\mu)
   \, \gamma_5 
   \, U_\mu^\dagger (x) \, \psi (x) \bigg) .
\label{eq:noether-axial}
\end{equation}
Below we list the results for the individual diagrams of the conserved
vector current. Due to chiral symmetry, the corresponding expressions
for the conserved axial current are trivially obtained by replacing
$\gamma_\mu$ with $\gamma_\mu\gamma_5$, and $\Gamma$ with
$\Gamma\gamma_5$. The vertex (diagram (a) in Fig. \ref{fig:diagrams})
gives the result
\begin{equation}
\frac{g_0^2}{16\pi^2} \,C_F \,\gamma_\mu\,\bigg[ -\log a^2p^2 +0.61800
+(1-\alpha) \Big(\log a^2p^2 -1.73375 \Big) \bigg] 
+c_{\rm vc}^{\rm vtx}(g_0^2) \cdot \Gamma,
\end{equation}
where the mixing coefficient $c_{\rm vc}^{\rm vtx}$ is given by
\begin{equation}
c_{\rm vc}^{\rm vtx}(g_0^2) \, = -0.43749 \cdot \frac{g_0^2}{16\pi^2}
\,C_F . 
\end{equation}
The result for the sails (diagrams (b) and (c) in
Fig.\,\ref{fig:diagrams}) is 
\begin{equation}
\frac{g_0^2}{16\pi^2} \,C_F \,\gamma_\mu\,\bigg[ 4.80841
-6.11653 \, (1-\alpha) \bigg] 
+c_{\rm vc}^{\rm sls}(g_0^2) \cdot \Gamma,
\end{equation}
where $c_{\rm vc}^{\rm sls}$ is obtained as
\begin{equation}
c_{\rm vc}^{\rm sls}(g_0^2) \, = -1.09017 \cdot \frac{g_0^2}{16\pi^2}
\,C_F . 
\end{equation}
Finally, the operator tadpole (diagram (d) in Fig.\,\ref{fig:diagrams})
gives the same result as for Wilson fermions:
\begin{equation}
 -g_0^2 \,C_F \,\gamma_\mu\, \frac{Z_0}{2}\, 
\Big(1-\frac{1}{4}(1-\alpha)\Big) .
\end{equation}
Summing up all contributions gives
\begin{equation}
\frac{g_0^2}{16\pi^2} \,C_F \,\gamma_\mu\,\bigg[ -\log a^2p^2 -6.80664
+(1-\alpha) \Big(\log a^2p^2 -4.79202 \Big) \bigg] 
+c_{\rm vc}(g_0^2) \cdot \Gamma,
\end{equation}
where the total mixing coefficient is given by
\begin{equation}
c_{\rm vc}(g_0^2) \, = -1.52766 \cdot \frac{g_0^2}{16\pi^2} \,C_F .
\end{equation}
These numbers exactly compensate the contributions of $\Sigma_1(p)$
and $c_1$ of eqs.\,(\ref{eq:Sigma1self}) and\,(\ref{eq:c1self}) of the
quark self-energy. Although we are not yet able to give an algebraic
proof that the term $i{\half}c_1\sum_{\mu\neq\nu} \gamma_\mu p_\nu$ in
the denominator of the self-energy, eq.\,(\ref{eq:renprop}), cancels
the unwanted contribution of the conserved currents proportional to
$\Gamma$, we know that this has to happen, for otherwise the
renormalization factors of the conserved currents would be different
from one. We have explicitly derived these currents using chiral Ward
identities, corresponding to transformations which leave the
Lagrangian invariant, and this proves that these currents are
conserved. Thus, it is certain that such cancellation must occur, and
the one-loop result $c_{\rm vc}=-c_1$ (which holds to all significant
digits that we have achieved) lends support to this statement.

It is for the above reasons why the term proportional to $c_1$ cannot
be absorbed into the redefinition of the four-momentum, as this would
spoil the correct renormalization of the conserved currents. Our
one-loop calculation has thus confirmed within our numerical precision
that the renormalization factors of these currents is unity, as
expected. It is remarkable that the use of the conserved currents
exactly cancels not only the self-energy terms that contribute to the
multiplicative renormalization, but also the mixing with
dimension-four operators, coming from hyper-cubic symmetry
breaking. Of course, radiative corrections to quark bilinears cannot
generate any terms proportional to $1/a$, and so an uncancelled,
power-divergent factor arising from the self-energy remains in the
total renormalization constant. However, as already stated, the latter
can be absorbed into a redefinition of the renormalized
four-momentum. By contrast, for the local vector and axial-vector
currents any mixing coming from the hyper-cubic breaking remains
uncancelled after the self-energy has been added to the vertex
diagram.

\bigskip
\bigskip

\par\noindent
{\bf 7.} In this article we have presented the first perturbative
study of a particular realization of minimally doubled fermions at
one-loop order. Our analysis has shown that minimally doubled fermions
are described at one loop by a fully consistent quantum field
theory. We have elucidated the consequences of the breaking of
hyper-cubic symmetry. In particular, we found that in the
Bori\c{c}i-Creutz construction\,\cite{mind:Borici07,mind:Creutz08},
all components of the four-momentum $p_\mu$ undergo a subtraction
under renormalization, which consists of a uniform shift.

Furthermore, local vector and axial-vector currents can mix with other
operators of the same dimension which are not invariant under the
hyper-cubic group. By contrast, no such mixing occurs for the scalar
density and the tensor current. We have derived expressions for the
conserved isospin-singlet vector and axial currents, which involve
only nearest-neighbour points. They do not undergo any mixing, and we
have verified that their renormalization constants are one. In fact,
apart from the staggered formulation, minimally doubled fermions are
the only discretization which yields a simple expression for a
conserved (point-split) axial-vector current.

It remains to discuss the implications of our findings for practical
simulations. Since there is one exact chiral $\rm U(1)\otimes U(1)$
symmetry, there must be exactly one Goldstone boson as the quark mass
is tuned to zero. It is natural to associate the neutral pion with
this particle. The charged pions, by contrast, will retain a finite
mass in the chiral limit at non-zero lattice spacing. An interesting
observation is that the influence of disconnected diagrams, which are
required for the determination of the $\pi^0$ mass, must become weaker
as the lattice spacing goes to zero, since the exact masslessness of
the charged pions is recovered in the continuum limit.\footnote{We
thank Mike Creutz for clarifying this issue.}

Even though the neutral pion in the chiral limit is exactly massless
in this theory, the renormalization of the four-momentum will modify
the rate of the exponential fall-off of its two-point correlation
function. Then, the extracted energies will be different for every
hadron from the ones given by the dispersion relations, and in
particular the rest energy of the neutral pion will not be zero (in
the chiral limit).

We can infer from our perturbative calculations,
eq.\,(\ref{eq:p-redefinition}), that this renormalization is governed
by one parameter, $c_2(g_0^2)$, with a further
explicit dependence on $a$. It remains to be investigated whether this
functional form is preserved at higher loop order and also
non-perturbatively, and what will be the practical prescriptions that
one has to infer from it. What is clear is that in numerical
simulations the subtraction would depend on $\beta$ and $a$ in a
different way from what the Callan-Symanzik renormalization group
equations dictate in the scaling region. This derives from the further
explicit dependence on the lattice spacing of the term multiplying
$c_2$.

We end our conclusions with a discussion of a possible strategy to
determine the renormalized momentum non-perturbatively. If we denote
the unsubtracted momentum by $p_\mu$, then the $\pi^0$ correlation
function at large Euclidean times will be
\be
   \sum_{\xvec}\rme^{i\pvec\cdot\xvec} \Big\langle
   \pi^0(x)\pi^{0\dagger}(0) \Big\rangle
   \propto A\rme^{-p_0x_0},\qquad {x_0\to\infty},
\ee
where $p_0$ is a function of the bare quark mass, $m_0$, and the
injected three-momentum $\pvec$. The relation to the temporal
component of the renormalized momentum, $p_{\mu;\rm ren}$, is then
given by
\be
   p_0(\pvec;m_0) = p_{0;\rm ren}(\pvec,p_\mu^{\rm cr};m_0),
\ee
where $p_\mu^{\rm cr}$ denotes the value of the four-momentum for
which the energies of the hadrons are restored to their physical
values. Since the $\pi^0$ is exactly massless in the chiral limit, the
above relation can serve as a renormalization condition to fix the
value of $p_\mu^{\rm cr}$. To this end, one has to evaluate
$p_0(\pvec;m_0)$ for several values of $m_0$ and $\pvec$ and determine
$p_\mu^{\rm cr}$ by implicitly solving the relation
\be
   \lim_{m_0\to0} p_0(\pvec;m_0)\Big|_{\pvec=\pvec^{\rm cr}} =
   p_0^{\rm cr}.
\ee
The energies in all hadronic channels can then be computed by setting
the momenta to $p_0^{\rm cr}$. The technical difficulty associated
with this strategy will be to control the large statistical
fluctuations in the $\pi^0$ channel which arise from disconnected
diagrams.

One may wonder if other variants of minimally doubled fermions, such
as the proposals by Karsten\,\cite{mind:Karsten81} and
Wilczek\,\cite{mind:Wilczek87} are easier to implement in
simulations. We are currently investigating this issue, but leave a
more thorough discussion to a future
publication\,\cite{workinprogress}.

\bigskip
\par\noindent
{\bf Acknowledgments} We thank Mike Creutz for useful discussions and
comments on the manuscript. This work was supported by Deutsche
Forschungsgemeinschaft (SFB443) and Gesellschaft f\"ur
Schwerionenforschung, GSI.


\bibliographystyle{h-elsevier}   
\bibliography{biblist}           

\end{document}